\documentclass[%
aps,prb
amsmath,amssymb,
reprint
]{revtex4-2}

\usepackage{dcolumn}
\usepackage{bm}
\usepackage{color}

\usepackage{ulem}

\usepackage{comment}
\usepackage[pdftex]{graphicx}
\usepackage{here}
\usepackage[version=3]{mhchem}
\usepackage{subfiles}
\usepackage{booktabs}
\usepackage{multirow}

\begin{document}

\title{Sublattice-selective inverse Faraday effect in ferrimagnetic rare-earth iron garnet}

\author{Toshiki Hiraoka}
 \affiliation{Department of Physics, Tokyo Institute of Technology, Meguro, Tokyo 152-8551, Japan}
\author{Ryo Kainuma}
\affiliation{Department of Physics, Tokyo Institute of Technology, Meguro, Tokyo 152-8551, Japan}
\author{Keita Matsumoto}
\affiliation{Department of Physics, Tokyo Institute of Technology, Meguro, Tokyo 152-8551, Japan}
\author{Kihiro T. Yamada}
\affiliation{Department of Physics, Tokyo Institute of Technology, Meguro, Tokyo 152-8551, Japan}
\author{Takuya Satoh}
 \email{satoh@phys.titech.ac.jp}
\affiliation{Department of Physics, Tokyo Institute of Technology, Meguro, Tokyo 152-8551, Japan}
\affiliation{Quantum Research Center for Chirality, Institute for Molecular Science, Okazaki, Aichi 444-8585, Japan}
\date{\today}

\begin{abstract}
We performed time-resolved pump--probe measurements using rare-earth iron garnet \ce{Gd3/2Yb1/2BiFe5O12} as a two-sublattice ferrimagnet.
We measured the initial phases of the magnetic resonance modes below and above the magnetization compensation temperature to clarify the sublattice selectivity of the inverse Faraday effect in ferrimagnets.
A comparison of the time evolution of magnetization estimated using the equations of motion revealed that the inverse Faraday effect occurring in ferrimagnetic materials has sublattice selectivity.
This is in striking contrast to antiferromagnets, in which the inverse Faraday effect acts on each sublattice identically.
The initial phase analysis can be applied to other ferrimagnets with compensation temperatures.
\end{abstract}

\maketitle

The ultrafast control of magnetic materials using light pulses has attracted considerable interest over the years.
In magnetic insulators, the inverse Faraday effect (IFE) and Faraday effect (FE) have been applied to excite and detect magnetization dynamics, respectively.\cite{Kimel05}
The IFE, which is the reverse of the magneto-optical FE, was theoretically proposed by Pitaevskii\cite{Pitaevskii61} and Pershan \cite{Pershan63} and demonstrated by van der Ziel et al.\cite{VanderZiel65}
Ultrafast magnetization control has been demonstrated in weak ferromagnets, such as \ce{DyFeO3}\cite{Kimel05} and \ce{FeBO3},\cite{Kalashnikova07} and pure antiferromagnets, such as NiO,\cite{Satoh10} via the IFE using femtosecond light pulses.
Here, an effective magnetic field pulse $\mathbf{H}_\mathrm{IFE}$ was induced by a circularly polarized light pulse.
The direction of $\mathbf{H}_\mathrm{IFE}$ is determined by the helicity of the circularly polarized light.
The impulsive $\mathbf{H}_\mathrm{IFE}$ acts on the sublattice magnetization of transition metal ions (e.g., Fe$^{3+}$ ions in \ce{DyFeO3} and \ce{FeBO3} and Ni$^{2+}$ ions in NiO), followed by precession
 as free induction decay.
In this theory, the magnitude and direction of $\mathbf{H}_\mathrm{IFE}$ acting on each sublattice are identical when $\mathbf{H}_\mathrm{IFE}$ is perpendicular to the sublattice magnetizations.\cite{Tzschaschel17}

Ferrimagnets consist of magnetic sublattices aligned in opposite directions with different magnetic ions of different magnitudes, such that the vector sum is not zero.
Thus, ferrimagnets exhibit a net magnetization in the ground state. Strong antiferromagnetic exchange interactions occur between the two magnetic sublattices.
Accordingly, the magnetic resonance in ferrimagnets has two modes: the ferromagnetic resonance (FMR) mode (GHz range), which is similar to that in ferromagnets, and the exchange resonance mode (sub-THz range), which is unique to ferrimagnets.\cite{Kaplan53,Wangsness53,Lax62}
Furthermore, when the temperature dependence of the sublattice magnetization differs, the net magnetization can vanish at the magnetization compensation temperature $T_\text{M}$.
Magnetization dynamics in ferrimagnets have been actively studied in the field of ferrimagnetic spintronics, which combines controllability, similar to ferromagnetism, with ultrafastness, similar to antiferromagnetism.\cite{Ivanov19,Finley20,Kim22}
	
\begin{figure}[t]
      \includegraphics[width=0.8\linewidth]{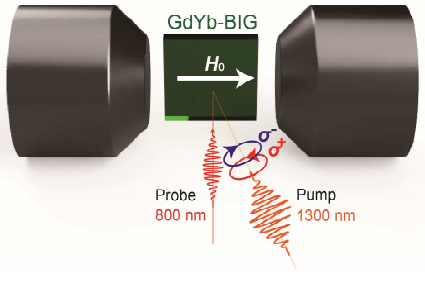}
  \caption{(Color online) Experimental geometry. The GdYb-BIG single crystal was placed in an electromagnet, which produces an in-plane magnetic field $\pm H_0$.} \label{setup}
\end{figure}	
	
Ultrafast IFE using circularly polarized light pulses has been reported in ferrimagnets.\cite{Stanciu07,Reid10,Satoh12,Parchenko13,Parchenko16,Deb16,Stupakiewicz21}
However, the magnetic field and temperature dependence of the initial phase were not explicitly discussed; moreover, whether the IFE generated $\mathbf{H}_\mathrm{IFE}$ in the opposite or same direction for each sublattice has not been studied.
We attempt to clarify this issue by measuring the initial phase of the coherently excited resonance modes using a two-sublattice rare-earth (RE) iron garnet.

The bismuth-doped RE iron garnet \ce{Gd3/2Yb1/2BiFe5O12} (GdYb-BIG) is a ferrimagnet with a Curie temperature of 573 K and $T_\text{M} = 96$~K, which was obtained from the temperature dependence of magnetization and the sign change of Faraday rotation.\cite{Parchenko14}
The exchange interaction between Fe ions in the tetragonal and octahedral sites is much stronger than that between Fe and RE ions.
Therefore, in the sub-THz frequency range, one can regard the two magnetic sublattices to be composed of Fe magnetization $\mathbf{M}^\text{Fe}$ and RE magnetization $\mathbf{M}^\text{RE}$ (Ref. \cite{Stupakiewicz21}).

We used a GdYb-BIG single crystal with a (111) plane orientation and a thickness of 140~$\mu$m, which was grown by the liquid-phase epitaxy method. 
As shown in Fig. \ref{setup}, we performed time-resolved pump--probe measurements in transmission geometry.
The polarization of pump pulse (wavelength 1300 nm, pulse energy 4 $\mu$J, repetition rate 500 Hz, pulse width 60 fs, spot diameter 280 $\mu$m) was circular ($\sigma^{\pm}$) to excite the magnetic resonance modes via the IFE.
The Faraday rotation angle of the linearly polarized probe pulse (wavelength 800 nm, pulse energy less than 10 nJ, repetition rate 1 kHz, pulse width 60 fs, spot diameter 80 $\mu$m) is mainly sensitive to the out-of-plane component $M_z^\text{Fe}$ over the entire temperature range.\cite{Khorsand13,Parchenko14}
We applied an external in-plane magnetic field $H_0 = 2$~kOe in the positive and negative directions, which was sufficient to align $\mathbf{M}^\text{Fe}$ and $\mathbf{M}^\text{RE}$ in the plane, resulting in a single-domain structure below 75 K and above 130 K.
Thus, we measured the dependences of the magnetization dynamics on the helicity ($\sigma^{\pm}$) of the circular pump polarization, the direction of the external magnetic field ($\pm H_0$), and the temperature ($T \gtrless T_\text{M}$).
The direction of $\mathbf{H}_\mathrm{IFE}$ can be controlled by $\sigma^{\pm}$; the directions of $\mathbf{M}^\text{Fe}$ and $\mathbf{M}^\text{RE}$ can be controlled by $\pm H_0$; and the relative magnitudes of $\mathbf{M}^\text{Fe}$ and $\mathbf{M}^\text{RE}$ can be controlled by $T$.


\begin{figure}[btp]
\includegraphics[width=0.9\linewidth]{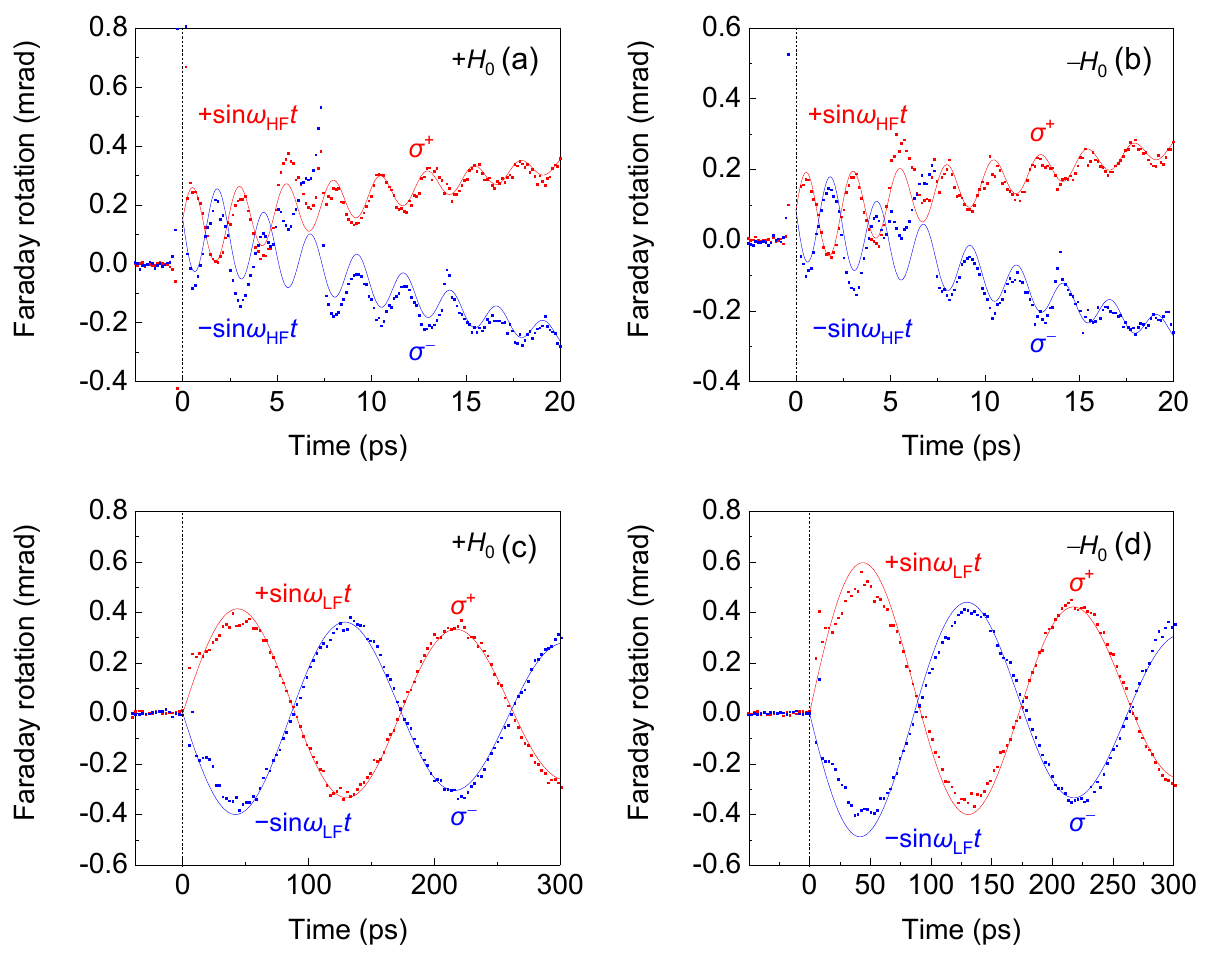}
\caption{(Color online) Magnetization dynamics at 300~K. HF mode at external field: $+H_0$ (a) and $-H_0$ (b). LF mode at external field: $+H_0$ (c) and $-H_0$ (d). Pump helicity: $\sigma^+$ (red dots) and $\sigma^-$ (blue dots). The solid lines are the damped sinusoidal functions with $\alpha_\mathrm{HF} \sim 0.03$ and $\alpha_\mathrm{LF} \sim 0.05$ for the HF and LF modes, respectively.} \label{dynamics_300K}
\end{figure}
\begin{figure}[btp]
      \includegraphics[width=0.9\linewidth]{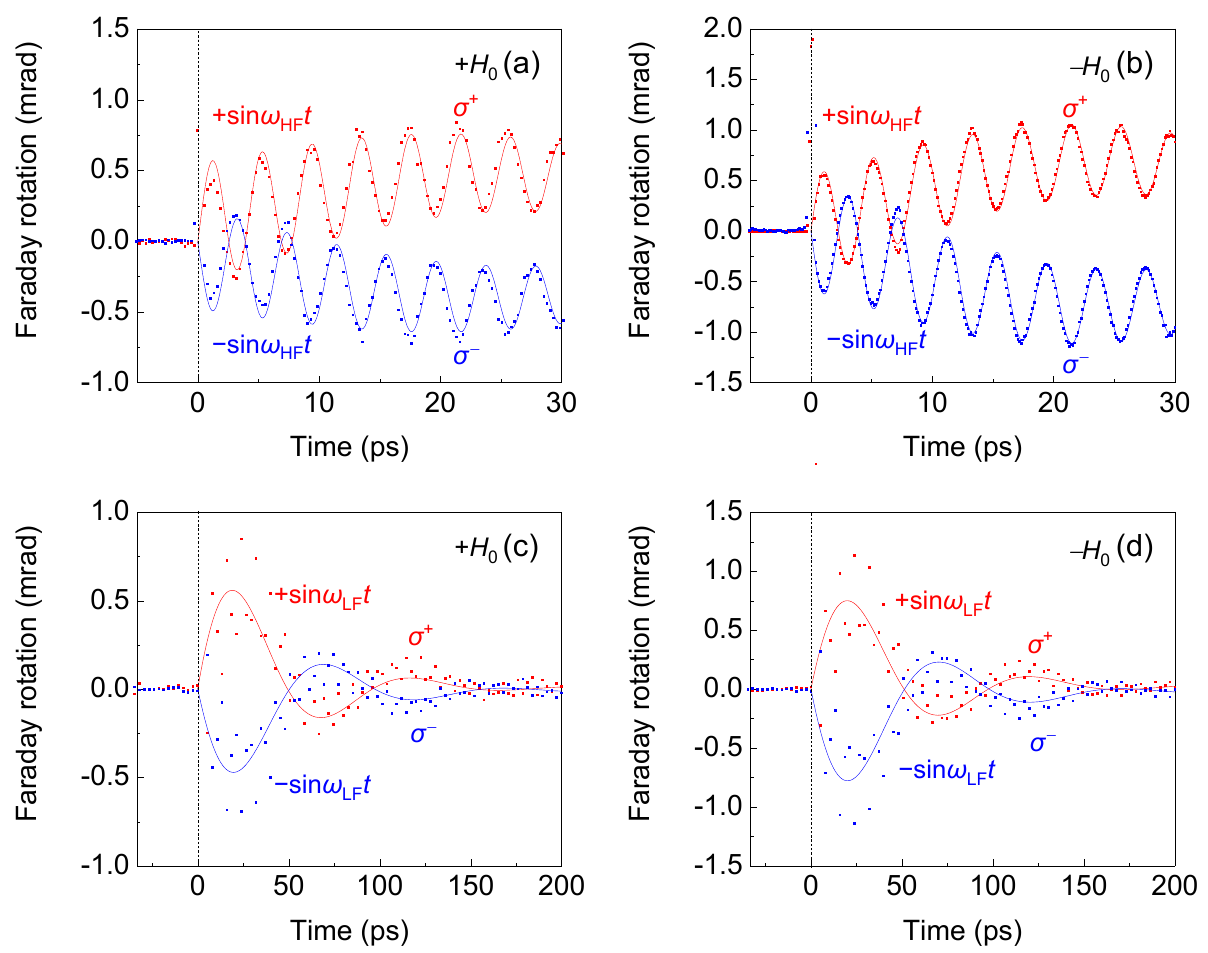}
  \caption{(Color online) Magnetization dynamics at 60~K. HF mode at external field: $+H_0$ (a) and $-H_0$ (b). LF mode at external field: $+H_0$ (c) and $-H_0$ (d). Pump helicity: $\sigma^+$ (red dots) and $\sigma^-$ (blue dots). The solid lines are the damped sinusoidal functions with $\alpha_\mathrm{HF} \sim 0.01$ and $\alpha_\mathrm{LF} \sim 0.35$ for the HF and LF modes, respectively.} \label{dynamics_60K}
\end{figure}

Figures \ref{dynamics_300K}(a)--\ref{dynamics_300K}(d) and \ref{dynamics_60K}(a)--\ref{dynamics_60K}(d) show the Faraday rotation of the probe as a function of delay $t$ at $T=300$~K ($>T_\text{M}$) and 60~K ($<T_\text{M}$), respectively, in the external magnetic fields $+H_0$ [(a): $t \le 20$ ps, (c): $t \le 300$ ps] and $-H_0$ [(b): $t \le 20$ ps, (d): $t \le 300$ ps], together with fitting with damped sinusoidal functions $\sin(\omega_\mathrm{HF} t)\exp(-\alpha_\mathrm{HF} \omega_\mathrm{HF} t)$ and $\sin(\omega_\mathrm{LF} t)\exp(-\alpha_\mathrm{LF} \omega_\mathrm{LF} t)$.
At 300 K, a high-frequency (HF) mode at 403 GHz and a low-frequency (LF) mode at 5.8 GHz were observed. 
At 60 K, an HF mode at 222 GHz and an LF mode at 10.7 GHz were observed.
The HF and LF modes were attributed to the exchange resonance and spatially propagating FMR (magnetostatic) modes, respectively.\cite{Parchenko13}
From Figs. \ref{dynamics_300K}(a)--\ref{dynamics_300K}(d) and \ref{dynamics_60K}(a)--\ref{dynamics_60K}(d), we observe that the initial phases of the sinusoidal functions [$\sin(\omega t)$ or $-\sin(\omega t)$] of the LF and HF modes do not depend on the temperature or the direction of the external field.
In contrast, the pump helicity changed the phases of both modes by 180$^{\circ}$.
The results are summarized in Table \ref{initial_phase_prediction_opposite}.
The peak at approximately 6--7 ps in Figs. \ref{dynamics_300K}(a) and \ref{dynamics_300K}(b) is due to the reflection of the pump pulse from the second face of the sample.\cite{Khan19}

It should be noted that for $T<T_\text{M}$, measurements were made at 40 and 50 K in addition to 60 K, with qualitatively the same results.
For $T<40$ K, the analysis is complicated by the contribution of Yb ions.
For $T_\text{M}<T$, measurements were performed at 140--300 K and similar results were obtained.
For $75~\text{K}<T<130~\text{K}$, the applied field was not sufficient to align the magnetization in-plane, and the two modes could not be excited.

\begin{table}[tb]
\caption{Time evolutions of the HF and LF modes (measurement and Case 1)}
\label{initial_phase_prediction_opposite}
\centering
\begin{tabular}{r}
\begin{tabular}{ccccc}
\toprule
Temperature & \multicolumn{4}{c}{$T>T_\mathrm{M}$} \\
Pump helicity & \multicolumn{2}{c}{$\sigma^+$} & \multicolumn{2}{c}{$\sigma^-$} \\
External field & $+H_0$ & $-H_0$ & $+H_0$ & $-H_0$ \\
\midrule
LF mode & $+\sin\omega_\mathrm{LF}t$ & $+\sin\omega_\mathrm{LF}t$ & $-\sin\omega_\mathrm{LF}t$ & $-\sin\omega_\mathrm{LF}t$ \\
HF mode & $+\sin\omega_\mathrm{HF}t$ & $+\sin\omega_\mathrm{HF}t$ & $-\sin\omega_\mathrm{HF}t$ & $-\sin\omega_\mathrm{HF}t$ \\
\bottomrule
\end{tabular} \\
\\
\begin{tabular}{ccccc}
\toprule
& \multicolumn{4}{c}{$T<T_\mathrm{M}$} \\
& \multicolumn{2}{c}{$\sigma^+$} & \multicolumn{2}{c}{$\sigma^-$} \\
& $+H_0$ & $-H_0$ & $+H_0$ & $-H_0$ \\
\midrule
& $+\sin\omega_\mathrm{LF}t$ & $+\sin\omega_\mathrm{LF}t$ & $-\sin\omega_\mathrm{LF}t$ & $-\sin\omega_\mathrm{LF}t$ \\
& $+\sin\omega_\mathrm{HF}t$ & $+\sin\omega_\mathrm{HF}t$ & $-\sin\omega_\mathrm{HF}t$ & $-\sin\omega_\mathrm{HF}t$ \\
\bottomrule
\end{tabular} \\
\end{tabular}
\end{table}

\begin{figure}[tb]
\includegraphics[width=0.85\linewidth]{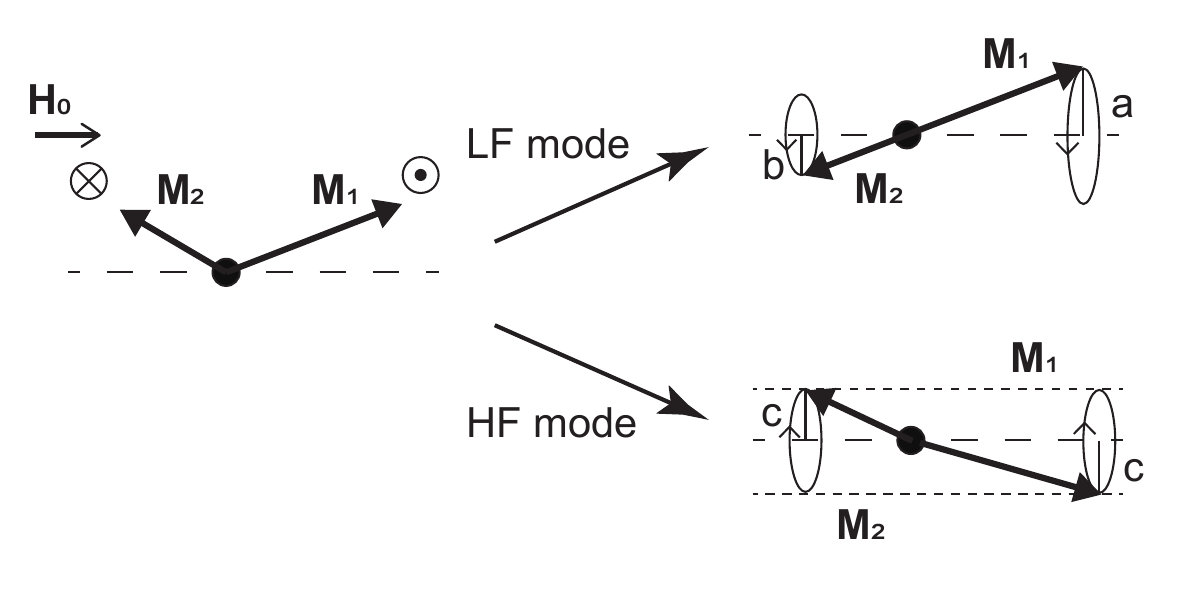}
\caption{Case 1: 
snapshot of sublattice magnetization deviations by the action of 
$\mathbf{H}_\mathrm{IFE}$ pulses with the opposite directions on each magnetic sublattice.
The cross and dot circles indicate that the directions of $\mathbf{H}_\mathrm{IFE}$ are from the front to the back and from the back to the front of the plane, respectively.
These deviations can be decomposed into those of the LF and HF modes.
$\mathbf{M}_\mathrm{1}$ and $\mathbf{M}_\mathrm{2}$ rotate counterclockwise and clockwise  around the $H_0$ axis in LF and HF modes, respectively.
}
\label{opposite_ife_dynamics}
\end{figure}

\begin{figure}[tb]
\includegraphics[width=0.85\linewidth]{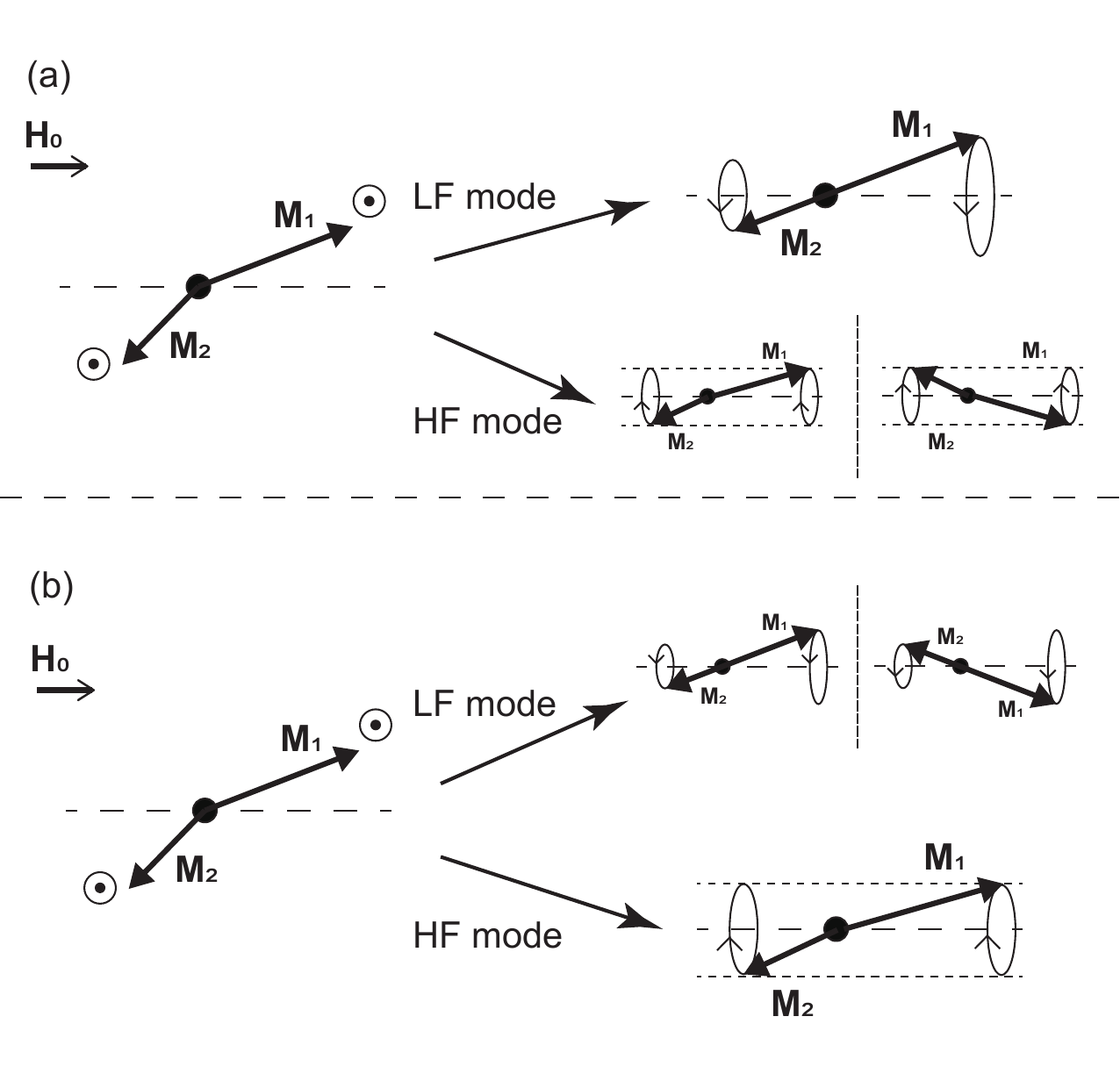}
\caption{Case 2: 
snapshot of sublattice magnetization deviations by the action of 
$\mathbf{H}_\mathrm{IFE}$ pulses with the same directions on each magnetic sublattice.
The dot circles indicate that the directions of $\mathbf{H}_\mathrm{IFE}$ are from the back to the front of the plane.
These deviations can be decomposed into those of the LF and HF modes. The dominant motion is in the LF mode (a) or HF mode (b).
$\mathbf{M}_\mathrm{1}$ and $\mathbf{M}_\mathrm{2}$ rotate counterclockwise and clockwise  around the $H_0$ axis in LF and HF modes, respectively.
}
\label{same_ife_dynamics}
\end{figure}

To understand the initial phases, two cases can be considered for the sublattice selectivity of the IFE.
The directions of $\mathbf{H}_\mathrm{IFE}^\text{Fe}$ acting on $\mathbf{M}^\text{Fe}$ and $\mathbf{H}_\mathrm{IFE}^\text{RE}$ acting on $\mathbf{M}^\text{RE}$ are opposite (Case 1 in Fig. \ref{opposite_ife_dynamics}) and the same (Case 2 in Fig. \ref{same_ife_dynamics}), respectively. 
In each case, $\mathbf{M}^\text{Fe}$ and $\mathbf{M}^\text{RE}$ rotate instantaneously according to the following equations of motion
under the impulsive actions of $\mathbf{H}_\mathrm{IFE}^\text{Fe}$ and $\mathbf{H}_\mathrm{IFE}^\text{RE}$, respectively.
\begin{align}
\frac{d\mathbf{M}^\text{Fe}}{dt}&=-\gamma \mathbf{M}^\text{Fe} \times \mathbf{H}_\mathrm{IFE}^\text{Fe}, \\
\frac{d\mathbf{M}^\text{RE}}{dt}&=-\gamma \mathbf{M}^\text{RE} \times \mathbf{H}_\mathrm{IFE}^\text{RE}.
\end{align}
Here, $\gamma$ is the gyromagnetic ratio, which is equal for Gd and Fe ions. Even if they were different, the qualitative argument in this paper would still hold.
Equations (1) and (2) are valid only during laser-pulse excitation, where the interactions between the sublattices, magnetic anisotropy field, external magnetic field, and damping can be neglected.\cite{Yoshimine14}
After the pulses of $\mathbf{H}_\mathrm{IFE}^\text{Fe}$ and $\mathbf{H}_\mathrm{IFE}^\text{RE}$ disappear, $\mathbf{M}^\text{Fe}$ and $\mathbf{M}^\text{RE}$ continue to rotate as a superposition of the counterclockwise LF and clockwise HF modes around the $H_0$ axis, which determine the initial phases of the two modes in Figs. \ref{dynamics_300K}(a)--\ref{dynamics_300K}(d) and \ref{dynamics_60K}(a)--\ref{dynamics_60K}(d).
Let the two sublattice magnetizations be $\mathbf{M}_1$ and $\mathbf{M}_2$, where $\left|\mathbf{M}_1\right|>\left|\mathbf{M}_2\right|$.
$\mathbf{M}_1$ points toward the external magnetic field.
At 300 K, $\mathbf{M}_1 = \mathbf{M}^\text{Fe}$ and $\mathbf{M}_2 = \mathbf{M}^\text{RE}$. At 60 K, $\mathbf{M}_1=\mathbf{M}^\text{RE}$ and $\mathbf{M}_2 = \mathbf{M}^\text{Fe}$.

In Case 1, let $a$ and $b$ be the in-plane displacements of $\mathbf{M}_1$ and $\mathbf{M}_2$ in the LF mode, respectively.
The ratio $a/b$ can be approximated as $\left|\mathbf{M}_1\right|/\left|\mathbf{M}_2\right|$.
In the HF mode, the in-plane displacements of $\mathbf{M}_1$ and $\mathbf{M}_2$ are regarded as identical and are denoted by $c$.\cite{Lax62}
The in-plane displacements of $\mathbf{M}_1$ and $\mathbf{M}_2$ can be represented by the superposition of the LF and HF modes if $a \geq c\geq b$ holds true.
In special cases where $\mathbf{H}_\mathrm{IFE}$ is generated only in $\mathbf{M}_1$ or $\mathbf{M}_2$, $a > b = c$ or $a = c > b$ hold true, respectively.
The dependences of the time evolution of $M_z^\text{Fe}$ on the pump helicity, the direction of the external magnetic field, and the temperature were consistent with the experimental results, as presented in Table \ref{initial_phase_prediction_opposite}.

In Case 2, the time evolution of $M_z^\text{Fe}$ differs depending on whether the dominant motion is in the LF mode (Case 2a) or HF mode (Case 2b), because the sense of rotation is opposite between the two modes.
In Case 2a [Fig. \ref{same_ife_dynamics}(a)], the $M_z^\text{Fe}$ for the LF mode can be determined, but not that for the HF mode.
In Case 2b [Fig. \ref{same_ife_dynamics}(b)], the $M_z^\text{Fe}$ for the HF mode can be determined, but not that for the LF mode.
The time evolutions of $M_z^\text{Fe}$ are presented in Tables \ref{initial_phase_prediction_same_fmr} and \ref{initial_phase_prediction_same_ex} for Cases 2a and 2b, respectively; however, they do not match the experimental results in Table \ref{initial_phase_prediction_opposite}.

\begin{table}[t]
\caption{Time evolutions of the HF and LF modes (Case 2a). The $\pm$ signs of the HF modes are in the same order.}
\label{initial_phase_prediction_same_fmr}
\centering
\begin{tabular}{r}
\begin{tabular}{ccccc}
\toprule
Temperature & \multicolumn{4}{c}{$T>T_\mathrm{M}$} \\
Pump helicity & \multicolumn{2}{c}{$\sigma^+$} & \multicolumn{2}{c}{$\sigma^-$} \\
External field & $+H_0$ & $-H_0$ & $+H_0$ & $-H_0$ \\
\midrule
LF mode & $+\sin\omega_\mathrm{LF}t$ & $+\sin\omega_\mathrm{LF}t$ & $-\sin\omega_\mathrm{LF}t$ & $-\sin\omega_\mathrm{LF}t$ \\
HF mode & $\pm\sin\omega_\mathrm{HF}t$ & $\pm\sin\omega_\mathrm{HF}t$ & $\mp\sin\omega_\mathrm{HF}t$ & $\mp\sin\omega_\mathrm{HF}t$ \\
\bottomrule
\end{tabular} \\
\\
\begin{tabular}{ccccc}
\toprule
& \multicolumn{4}{c}{$T<T_\mathrm{M}$} \\
& \multicolumn{2}{c}{$\sigma^+$} & \multicolumn{2}{c}{$\sigma^-$} \\
& $+H_0$ & $-H_0$ & $+H_0$ & $-H_0$ \\
\midrule
& $-\sin\omega_\mathrm{LF}t$ & $-\sin\omega_\mathrm{LF}t$ & $+\sin\omega_\mathrm{LF}t$ & $+\sin\omega_\mathrm{LF}t$ \\
& $\mp\sin\omega_\mathrm{HF}t$ & $\mp\sin\omega_\mathrm{HF}t$ & $\pm\sin\omega_\mathrm{HF}t$ & $\pm\sin\omega_\mathrm{HF}t$ \\
\bottomrule
\end{tabular} \\
\end{tabular}
\end{table}

\begin{table}[b]
\caption{Time evolutions of the HF and LF modes (Case 2b). The $\pm$ signs of the LF modes are in the same order.}
\label{initial_phase_prediction_same_ex}
\centering
\begin{tabular}{r}
\begin{tabular}{ccccc}
\toprule
Temperature & \multicolumn{4}{c}{$T>T_\mathrm{M}$} \\
Pump helicity & \multicolumn{2}{c}{$\sigma^+$} & \multicolumn{2}{c}{$\sigma^-$} \\
External field & $+H_0$ & $-H_0$ & $+H_0$ & $-H_0$ \\
\midrule
LF mode & $\pm\sin\omega_\mathrm{LF}t$ & $\pm\sin\omega_\mathrm{LF}t$ & $\mp\sin\omega_\mathrm{LF}t$ & $\mp\sin\omega_\mathrm{LF}t$ \\
HF mode & $+\sin\omega_\mathrm{HF}t$ & $+\sin\omega_\mathrm{HF}t$ & $-\sin\omega_\mathrm{HF}t$ & $-\sin\omega_\mathrm{HF}t$ \\
\bottomrule
\end{tabular} \\
\\
\begin{tabular}{ccccc}
\toprule
& \multicolumn{4}{c}{$T<T_\mathrm{M}$} \\
& \multicolumn{2}{c}{$\sigma^+$} & \multicolumn{2}{c}{$\sigma^-$} \\
& $+H_0$ & $-H_0$ & $+H_0$ & $-H_0$ \\
\midrule
& $\mp\sin\omega_\mathrm{LF}t$ & $\mp\sin\omega_\mathrm{LF}t$ & $\pm\sin\omega_\mathrm{LF}t$ & $\pm\sin\omega_\mathrm{LF}t$ \\
& $-\sin\omega_\mathrm{HF}t$ & $-\sin\omega_\mathrm{HF}t$ & $+\sin\omega_\mathrm{HF}t$ & $+\sin\omega_\mathrm{HF}t$ \\
\bottomrule
\end{tabular} \\
\end{tabular}
\end{table}

We conclude that the IFE in 
GdYb-BIG operates in opposite directions for each sublattice (Case 1).
In special cases, $\mathbf{H}_\mathrm{IFE}$ is generated only in $\mathbf{M}^\text{Fe}$ or $\mathbf{M}^\text{RE}$.
This is consistent with the sublattice selectivity of the FE, in which the signs of the contributions of Fe and RE ions to the FE are opposite and the Fe ion's contribution is dominant in the near-infrared range.\cite{Khorsand13}
This is in contrast to antiferromagnets, in which the magnitude and direction of $\mathbf{H}_\mathrm{IFE}$ acting on each sublattice are identical.\cite{Tzschaschel17}
The initial phase analysis can be applied to other ferrimagnets by measuring below and above the compensation temperatures.
A clarification of the sublattice selectivity of the IFE is expected to promote the development of devices that utilize ferrimagnetic materials and magnetooptical effects, such as the study of magnetization reversal using the IFE.\cite{Dannegger21}

\begin{acknowledgments}
We are grateful to Kouki Mikuni for his technical assistance.
This study was financially supported by the Japan Society for the
Promotion of Science KAKENHI (grant Nos. JP19H01828, JP19H05618,
JP21H01032, JP22H01154, and JP22K14588), 
the Frontier Photonic Sciences Project (NINS grant Nos. 01212002 and 01213004), 
and OML Project (NINS grant No. OML012301) of the National Institutes of Natural Sciences (NINS), 
and MEXT Initiative to Establish NeXt-generation Novel Integrated Circuits CenterS (X-NICS) (grant No. JPJ011438).

\end{acknowledgments}


\bibliographystyle{apsrev4-2}
\bibliography{reference}

\end{document}